\documentclass[reprint,superscriptaddress,
 amsmath,amssymb,
 aps,
pra,
]{revtex4-2}

\usepackage{graphicx}
\usepackage{dcolumn}
\usepackage{bm}
\usepackage{physics}
\usepackage{amsfonts, amsmath}
\usepackage{dsfont}
\usepackage{amsthm}
\usepackage{bm}
\usepackage{hyperref}
\usepackage{bbold}
\usepackage{color}
\usepackage{xcolor}
\usepackage{lipsum,babel}
\usepackage[normalem]{ulem}

\DeclareMathOperator*{\V}{\mathbb{V}}

\newcommand{\cor}[1]{{\color{red}{#1}}}

\newcommand{\circlenum}[1]{\raisebox{.5pt}{\textcircled{\raisebox{-.9pt} {#1}}}}

\begin{document}
\definecolor{navy}{RGB}{46,72,102}
\definecolor{pink}{RGB}{219,48,122}
\definecolor{grey}{RGB}{184,184,184}
\definecolor{yellow}{RGB}{255,192,0}
\definecolor{grey1}{RGB}{217,217,217}
\definecolor{grey2}{RGB}{166,166,166}
\definecolor{grey3}{RGB}{89,89,89}
\definecolor{red}{RGB}{255,0,0}

\preprint{APS/123-QED}

\title{Efficacy of virtual purification-based error mitigation on quantum metrology}
\author{Hyukgun Kwon}
\affiliation{Department of Physics and Astronomy, Seoul National University, Seoul 08826, Republic of Korea}
\author{Changhun Oh}
\affiliation{Pritzker School of Molecular Engineering, University of Chicago, Chicago, Illinois 60637, USA}
\author{Youngrong Lim}
\affiliation{School of Computational Sciences, Korea Institute for Advanced Study, Seoul 02455, Korea}
\author{Hyunseok Jeong}
\email{h.jeong37@gmail.com}
\affiliation{Department of Physics and Astronomy, Seoul National University, Seoul 08826, Republic of Korea}
\author{Liang Jiang}
\email{liangjiang@uchicago.edu}
\affiliation{Pritzker School of Molecular Engineering, University of Chicago, Chicago, Illinois 60637, USA}

\begin{abstract}
Noise is the main source that hinders us from fully exploiting quantum advantages in various quantum informational tasks.
However, characterizing and calibrating the effect of noise is not always feasible in practice.
Especially for quantum parameter estimation, an estimator constructed without precise knowledge of noise entails an inevitable bias.
Recently, virtual purification-based error mitigation (VPEM) has been proposed to apply for quantum metrology to reduce such a bias occurring from unknown noise. 
While it was demonstrated to work for particular cases, whether VPEM always reduces a bias for general estimation schemes is unclear yet. 
For more general applications of VPEM to quantum metrology, we study factors determining whether VPEM can reduce the bias. 
We find that the closeness between the dominant eigenvector of a noisy state and the ideal quantum probe (without noise) with respect to an observable determines the reducible amount of bias by VPEM.
Next, we show that one should carefully choose the reference point of the target parameter, which gives a smaller bias than others because the bias depends on the reference point. 
Otherwise, even if the dominant eigenvector and the ideal quantum probe are close, the bias of the mitigated case could be larger than the non-mitigated one. 
Finally, we analyze the error mitigation for a phase estimation scheme under various noises. 
Based on our analysis, we predict whether VPEM can effectively reduce a bias and numerically verify our results.
\end{abstract}
              
\maketitle

\section{\label{sec:level1}Introduction}

Quantum metrology is a study of exploiting quantum properties to surpass the classical limit of estimation precision \cite{giovannetti2001quantum,baumgratz2016quantum,giovannetti2006quantum,Giovannetti2011,braun2018quantum,Pirandola2018}.
For various cases, quantum resources such as entanglement or squeezing enable us to achieve a better estimation precision that cannot be attained by classical means.
Especially for a phase estimation scheme in an interferometer system, which is one of the most important tasks in quantum metrology, the quantum advantage using various nonclassical states, such as N00N state \cite{PhysRevA.54.R4649,lee2002quantum,Demkowicz-Dobrzanski2015} and squeezed state \cite{Demkowicz-Dobrzanski2015,caves1981quantum,PhysRevLett.100.073601,abadie2011gravitational,aasi2013enhanced}, have been studied and experimentally demonstrated \cite{nagata2007beating,mitchell2004super,yonezawa2012quantum,berni2015ab,yu2020quantum}.

To consider more practical situations, it is vital to study the effect of noise on estimation performance. Many studies investigated how the performance deteriorates due to noise, mostly focusing on statistical error \cite{escher2011general,PhysRevLett.106.153603}. 
In practice, however, it is not always feasible to obtain complete information about noise. 
Such a lack of knowledge of the noise in the system results in a bias of the estimator, which cannot be reduced simply by increasing the number of samples, unlike the statistical error. Especially, when there are sufficiently many samples for an estimation where the bias becomes dominant compared to the statistical error, the bias determines the order of magnitude of the total estimation error.
In addition, while quantum error correction can recover the noisy state to its original state~\cite{PhysRevLett.112.080801,PhysRevLett.112.150801,PhysRevLett.112.150802,PhysRevX.7.041009,zhou2018achieving,PhysRevLett.122.040502,zhuang2020distributed,PRXQuantum.2.010343}, it requires many resources, such as mid-circuit measurement, low-noise threshold and larger number of required qubits because of encoding, which may not be practically accessible in the current era.

To address the above issues, Ref.~\cite{yamamoto2021error} has recently proposed applying virtual-purification-based error mitigation (VPEM) \cite{PhysRevX.11.031057,PhysRevX.11.041036} to quantum metrology to suppress the bias occurring from unknown noise and
shown that it considerably reduces the bias of an estimator of the phase estimation with a GHZ state. 
The principle behind VPEM is that it purifies the output state, which has undergone an unknown noise, to its dominant eigenstate, so that it effectively enables one to exploit the dominant eigenstate instead of the noisy state \cite{PhysRevX.11.031057,PhysRevX.11.041036}.
While Ref.~\cite{yamamoto2021error} presented a framework of VPEM for quantum metrology and provided an example in which VPEM effectively reduces a bias,
the general applicability of VPEM to quantum metrology has not been fully understood. 

In this work, we study when VPEM can effectively reduce the bias from unknown noise. 
In particular, we identify two crucial factors that determine whether VPEM can reduce the bias. 
The first one is the closeness of the dominant eigenvector of a noisy state and the (noiseless) ideal state with respect to an observable. 
Because VPEM only allows us to exploit the dominant eigenvector, not necessarily the ideal state, the dominant eigenvector has to be guaranteed to be sufficiently close to the ideal state.
We find that the closeness between the expectation values of an observable $\hat{A}$ (used for estimation) of the ideal quantum probe and the dominant eigenvector of a noisy state, determines how much bias can be reduced by VPEM. 
More specifically, the closeness decides the leading order of the reduced bias in the noise strength.
In addition, we show that applying VPEM does not necessarily reduce the bias compared to without VPEM.
The second one is the reference point of a parameter that we want to estimate. 
Assuming that the unknown parameter to be estimated is small (often called local parameter estimation \cite{paris2009quantum, Morelli_2021}), the reference point is the one around which the unknown parameter varies.

In local parameter estimation, we show that one has to carefully choose the reference point that gives a smaller bias than others before applying VPEM to quantum parameter estimation. 
Otherwise, even if the expectation values of $\hat{A}$ over the ideal quantum probe and the dominant eigenvector are the same, the bias of the mitigated case could be larger than the non-mitigated case. 
We emphasize that considering a reference point is a unique feature of quantum metrology that has not been considered in the previous study \cite{yamamoto2021error}.

To elaborate on the effect of the two aforementioned factors on VPEM, we apply our analysis to the phase estimation with the parity measurement in an interferometer system and numerically verify our analysis. 
We consider a N00N state and product of coherent and squeezed state as quantum probes for the phase estimation in the presence of phase diffusion, photon loss, and additive Gaussian noise. 
For the N00N state with phase diffusion or photon loss and for the product of a coherent and squeezed state in the presence of a special type of additive Gaussian noise, we find that the bias can be reduced by VPEM because the ideal quantum probe and the dominant eigenvector of the noisy state are the same, which is a similar case to the previous study \cite{yamamoto2021error}.
In addition, we also find that the bias of the mitigation case could be larger than the non-mitigated case if one does not select a proper reference point.
In contrast, for the product of coherent and squeezed state in the presence of photon loss, we find that it cannot benefit from VPEM, i.e., the magnitude of bias of the non-mitigated case and mitigated case are similar, even if we adopt \textit{good} reference point that gives smaller bias than others because the expectation values of the parity over the ideal quantum probe and the dominant eigenvector are not close enough.
It indicates that VPEM does not always promise to reduce the bias error.

This paper is organized as follows: In Sec.~\ref{S2V}, we explain how a bias occurs when one disregards unknown noise, and introduce VPEM. In Sec.~\ref{S3V}, we show the reducible amount of bias through VPEM. In particular, in Sec.~\ref{S3AV}, we inspect the relation between the dominant eigenvector and the leading order of the bias with respect to noise strength. In Sec.~\ref{S3BV}, we show how the bias depends on the reference point, highlighting the importance of the reference point when one applies VPEM to parameter estimation. In Sec.~\ref{S4V}, we apply our analysis to the phase estimation in the interferometer system.

\section{MSE with bias} \label{S2V}
Before introducing our main results, let us explain a basic quantum estimation scheme and how a bias occurs in the presence of an unknown noise. 

\subsection{Ideal Case}
Let us consider a unitary process that encodes an unknown parameter $\phi$. 
To estimate the parameter $\phi$, a prepared quantum probe $\hat{\rho}^{(0)}_{\text{id}}=|\psi^{(0)}_{\text{id}}\rangle\langle \psi^{(0)}_{\text{id}}|$, assumed to be a pure state, encodes the parameter by a unitary operation $\hat{\mathcal{U}}(\phi+\phi_{0})$, which results in the ideal output state ${\hat{\rho}_{\text{id}}(\phi+\phi_{0})=\dyad{\psi_{\text{id}}(\phi+\phi_{0})}}$. 
Here, $\phi_{0}$ is a reference point of $\phi$ that one can freely select on purpose.  
One then measures the state in the eigenbasis of an observable $\hat{A}$ for $N_{s}$ times, which renders the outcomes $\{A_{\text{id},k}\}^{N_{s}}_{k=1}$, where $A_{\text{id},k}$ is the $k$th measurement outcome.
Meanwhile, the expectation value of $\hat{A}$ over $\hat{\rho}_{\text{id}}$ is written as
\begin{align}
    \Tr\left[\hat{A}\hat{\rho}_{\text{id}}(\phi+\phi_0) \right]
    &=\bra{\psi_{\text{id}}(\phi+\phi_0)}\hat{A}\ket{\psi_{\text{id}}(\phi+\phi_0)}  \\ 
    &=x_{\text{id}} + y_{\text{id}}\phi+O(\phi^{2}),
\end{align}
where the expectation value is linearized for small $\phi$ with the coefficients $x_\text{id}$ and $y_\text{id}$
\begin{align}
    x_\text{id}(\phi_{0}) \equiv \text{Tr}[\hat{A}\hat{\rho}_\text{id}(\phi_{0})],
    ~y_\text{id}(\phi_{0}) \equiv \pdv{\text{Tr}[\hat{A}\hat{\rho}_\text{id}(\phi_0+\phi)]}{\phi} \bigg \vert_{\phi=0}. \label{xydef}
\end{align}

\begin{figure*}[t]
    \centering
    \includegraphics[width=\textwidth]{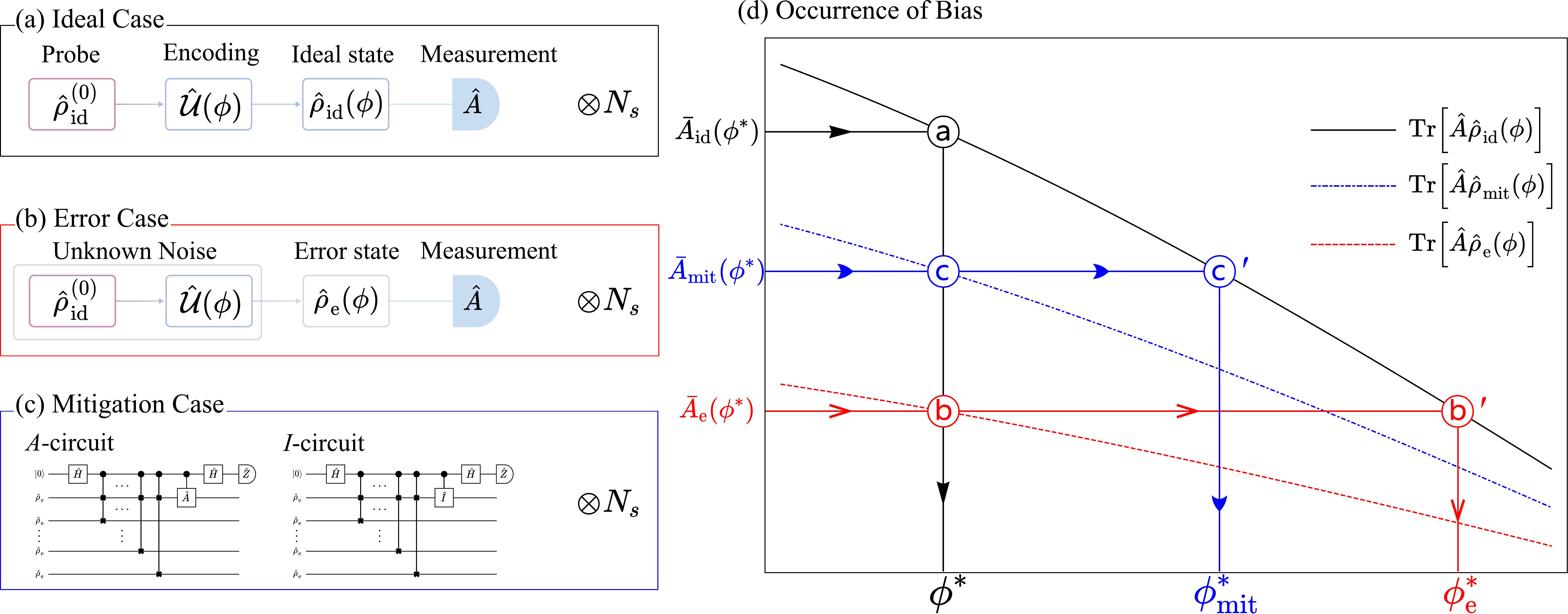}
    \caption{Schematic diagram of quantum metrology and the effect of VPEM. Here we ignore the statistical error to focus on the bias such that measurement outcomes $\bar{A}$ is equal to $\Tr\left[\hat{A}\hat{\rho}(\phi)\right]$ ($\bar{A}$ and $\hat{\rho}$ indicate $\hat{\rho}_{\text{id}/\text{e}/\text{mit}}$ and $\bar{A}_{\text{id}/\text{e}/\text{mit}}$) and set $\phi_{0}=0$ for simplicity. (a) The $\phi$ estimation process of the ideal case. Let us assume that encoded parameter $\phi$ is $\phi^{*}$ and one gets the average of measurement outcomes $\bar{A}_{\text{id}}(\phi^{*})$. One then compares the average with the theoretical value $\text{Tr}[\hat{A}\hat{\rho}_{\text{id}}(\phi^{*})]$ and estimate the encoded parameter as $\phi^{*}$; thus there is no bias. (See \textcircled{a} in (d).) 
    (b) The $\phi$ estimation process in the presence of an unknown noise which renders measurement outcomes $\bar{A}_{\text{e}}(\phi^{*})$. (See \textcircled{b} in (d).) 
    Since one still uses the ideal value $\text{Tr}[\hat{A}\hat{\rho}_{\text{id}}]$ for estimation, 
    one estimates $\phi$ as $\phi^{*}_{\text{e}}$ that satisfies $\bar{A}_{\text{e}}(\phi^{*})=\text{Tr}[\hat{A}\hat{\rho}_{\text{id}}(\phi^{*}_{\text{e}})]$, which results in a bias $\phi^{*}_{\text{e}}-\phi^{*}$. (See \textcircled{b}$'$ in (d).)    
    (c) VPEM effectively replaces the error state $\hat{\rho}_{\text{e}}(\phi)$ with $\hat{\rho}_{\text{mit}}(\phi)$. 
    We assume that the expectation value of $\hat{A}$ over the mitigated state and the ideal state is close enough, i.e., $\text{Tr}[\hat{A}\hat{\rho}_{\text{mit}}(\phi)]\approx \text{Tr}[\hat{A}\hat{\rho}_{\text{id}}(\phi)]$. 
    Similar to the error case, one estimates the encoded parameter as $\phi^{*}_{\text{mit}}$ whose bias is $\phi^{*}_{\text{mit}}-\phi^{*}$ which is smaller than the error case. Here $\phi^{*}_{\text{mit}}$ satisfies $\bar{A}_{\text{mit}}(\phi^{*})=\text{Tr}[\hat{A}\hat{\rho}_{\text{id}}(\phi^{*}_{\text{mit}})]$. (See \textcircled{c} and \textcircled{c}$'$ in (d).)
    }
    \label{fig:schematic}
\end{figure*}

Thus, the following estimator is unbiased for small $\phi$,
\begin{align}
    \phi^{\text{est}}_{\text{id}}=\frac{1}{y_{\text{id}}}\left(\bar{A}_{\text{id}}-x_{\text{id}}\right), \label{estimatorideal}
\end{align}
where $\bar{A}_{\text{id}}$ is the average of measurement outcomes $\sum_{k=1}^{N_{s}}{A_{\text{id},k}}/{N_{s}}$
(See Fig.~\ref{fig:schematic} for the schematic of the estimation.).

The mean squared error (MSE) of the estimator is given by
\begin{align}\label{esteideal}
    \delta^{2}\phi_{\text{id}}
    =\frac{1}{y_{\text{id}}^{2}}\frac{\V[\hat{A}]_{\hat{\rho}_{\text{id}}}}{N_{s}},
\end{align}
which depends only on the statistical error that originates from the fluctuation of the average of the measurement outcomes due to the finite sample size.
Here, $\V[\hat{A}]_{\hat{\rho}} \equiv \text{Tr}[\hat{A}^{2}\hat{\rho}]-\text{Tr}[\hat{A}\hat{\rho}]^{2}$ is the variance of $\hat{A}$ over the quantum state. 
Therefore, the MSE can be reduced by increasing the number of samples $N_{s}$. 
It is worth emphasizing that the convergence rate is closely related to Fisher information \cite{braunstein1994statistical, paris2009quantum}, which is the main focus of many studies about quantum metrology \cite{giovannetti2006quantum,Giovannetti2011}.
Our interest lies in the bias rather than the statistical error, which will be introduced in the following section.

\subsection{Error case}
In a more practical scenario where noise occurs during estimation, the ideal state $\hat{\rho}_{\text{id}}$ is replaced by an error state $\hat{\rho}_{\text{e}}$, which depends on the noise.
As the ideal case, one prepares $N_{s}$ numbers of $\hat{\rho}_{\text{e}}$ and obtains measurement outcomes $\{A_{\text{e},k}\}^{N_{s}}_{k=1}$, where $A_{\text{e},k}$ is the $k$th measurement outcome from $\hat{\rho}_{\text{e}}$. 
When one does not know that the noise occurs and exploits the same theoretical value $\text{Tr}[\hat{A}\hat{\rho}_{\text{id}}(\phi+\phi_{0})]$ as the ideal case, the estimator is then written as
\begin{align}
    \phi^{\text{est}}_{\text{e}}=\frac{1}{y_{\text{id}}}\left(\bar{A}_{\text{e}}-x_{\text{id}}\right),\label{estimatore}
\end{align}
where $\bar{A}_{\text{e}}$ is the average of measurement outcomes $\sum_{k=1}^{N_{s}}{A_{\text{e},k}}/{N_{s}}$.

Due to this choice, the estimator $\phi^{\text{est}}_{\text{e}}$ generally has a bias (See Fig.~\ref{fig:schematic}(d) for the schematic of how bias occurs.),
\begin{align}
    \left<\phi^{\text{est}}_{\text{e}}\right>-\phi 
    &= \frac{x_{\text{e}}-x_{\text{id}}+(y_{\text{e}}-y_{\text{id}})\phi }{y_{\text{id}}}+O(\phi^{2}) \label{estiavge}\\
    &\equiv B_{\text{e}}(\phi,\phi_{0},\Delta), \label{biase}
\end{align}
where $x_\text{e}$ and $y_\text{e}$ are defined by a similar relation to the ideal case
\begin{align}
     \Tr\left[\hat{A}\hat{\rho}_{\text{e}}(\phi+\phi_{0}) \right]
     &=  x_{\text{e}} + y_{\text{e}}\phi +O(\phi^{2}),  
\end{align}
and $\big<\cdots\big>$ is an average over all possible measurement outcomes of the quantum state. 
Suppose one can completely characterize the noise and $\hat{\rho}_\text{e}$, which is a typical assumption in many previous studies but may not always be feasible; one can then construct an unbiased estimator using $x_\text{e}$ and $y_\text{e}$ as the ideal case.
Thus, the bias occurs from the lack of information about the noise.
For our case, since $\phi^{\text{est}}_{\text{e}}$ is a biased estimator, the corresponding MSE $\delta^{2}\phi_{\text{e}}$, contains an additional term due to the bias:
\begin{align}
    \delta^{2}\phi_{\text{e}}  =\left(B_{\text{e}}\right)^{2}+\frac{1}{N_{s}}\frac{\V[\hat{A}]_{\hat{\rho}_{\text{e}}}}{y_{\text{id}}^{2}}. \label{esterre}  
\end{align}
For the rest of the paper, we call $\left(B_{\text{e}}\right)^{2}$ as a bias error of error case. Fig.~\ref{fig:schematic} presents the schematic of the occurrence of the bias. In contrast to the statistical error, one cannot reduce the bias by simply increasing the number of samples $N_{s}$. We emphasize that especially when there are sufficiently many samples (i.e., large $N_{s}$), the bias is more critical than the statistical error, in terms that the MSE is dominated by the bias.

\subsection{Error Mitigated Quantum Metrology}
\begin{figure}[t]
    \centering
    \includegraphics[width=0.95\linewidth]{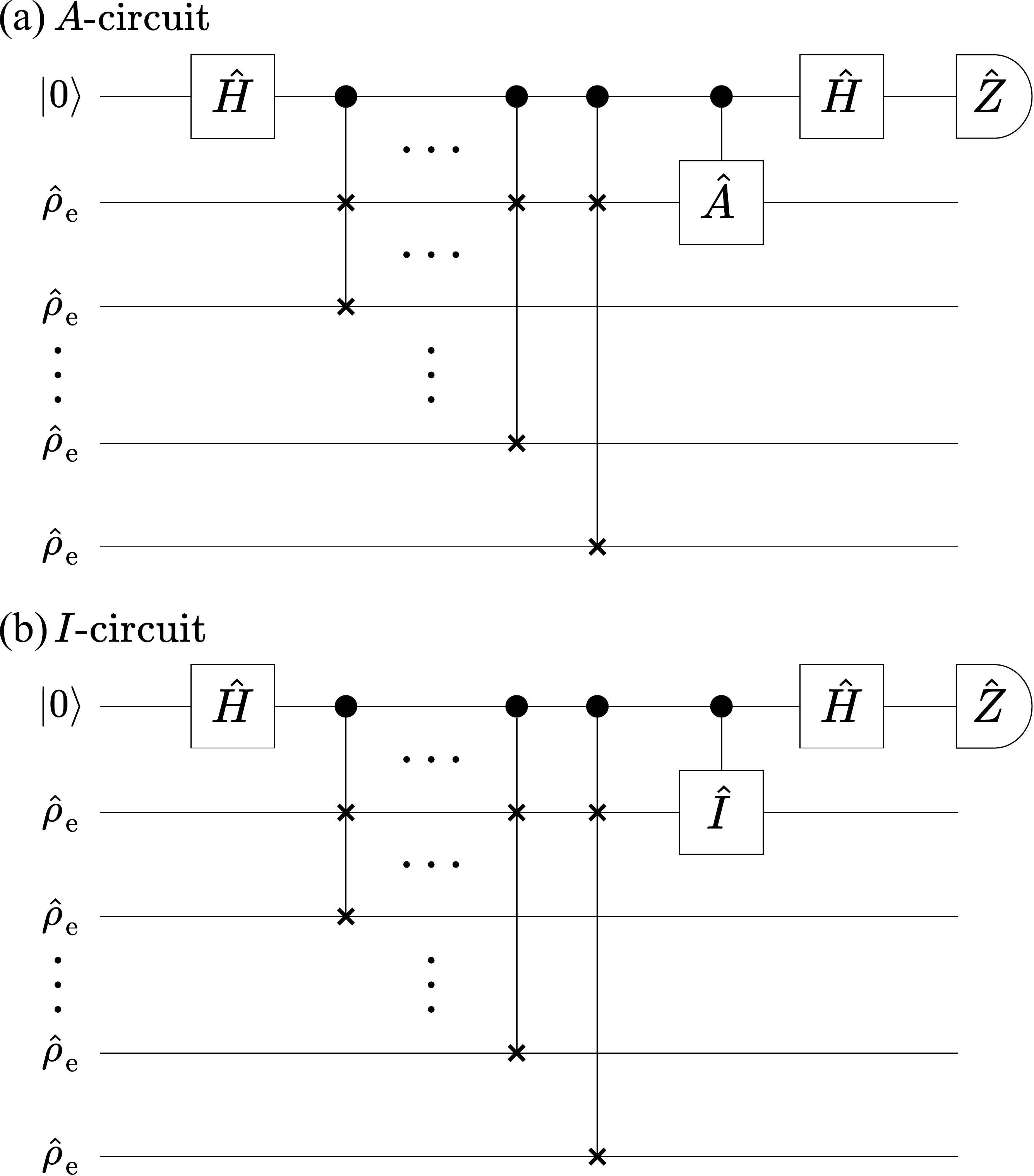}
    \caption{Error mitigation circuits. After a Hadamard gate on the ancilla qubit, one applies a sequence of $n$ controlled swap gates, followed by a controlled-$\hat{A}$ ($\hat{I}$) and another Hadamard gate, and measures in $\hat{Z}$-basis.}
    \label{fig:Circuits2}
\end{figure}

In this section, we introduce the scheme of VPEM quantum metrology, recently proposed in Ref.~\cite{yamamoto2021error}. 
The scheme employs $N_{s,\text{mit}}$ copies of so-called $A$-circuit and $I$-circuit, each of which exploits $n$ copies of an error state \cite{PhysRevX.11.031057} (see Fig.~\ref{fig:Circuits2}).
Although one might iterate each circuit for different numbers, we equally set $N_{s,\text{mit}}$ as the number of iterations for each circuit for simplicity. 
The details of the dynamics are described in Refs.~\cite{yamamoto2021error, PhysRevX.11.031057, PhysRevX.11.041036}. 

By implementing the circuits with an error state with its diagonal form being
\begin{align}
   \hat{\rho}_{\text{e}}(\phi+\phi_{0}) =\lambda \dyad{\psi} + (1-\lambda) \sum_{k=1} p_{k} \dyad{\psi_k}, \label{eq:diag}
\end{align}
one obtains measurement outcomes $\left\{z^{A}_{k}\right\}^{N_{s,\text{mit}}}_{k=1}$ and $\left\{z^{I}_{k}\right\}^{N_{s,\text{mit}}}_{k=1}$ from each circuit, where $z^{A}_{k}, z^{I}_{k}\in \{1,-1\}$ for all $k$'s.
Here, $\sum_{k=1} p_{k}=1$ and $\lambda$ is the largest eigenvalue and $|\psi\rangle$ is the corresponding eigenvector, which may be different from the ideal state $\ket{\psi_{\text{id}}}$. 
Also, the relevant parameters in Eq.~\eqref{eq:diag} are functions of $(\phi+\phi_{0})$ and noise strength $\Delta$.
After an experiment, we obtain the average values as
\begin{align}
    &\bar{Z}^{A} \equiv \frac{1}{N_{s,\text{mit}}}\sum^{N_{s,\text{mit}}}_{k=1}z^{A}_{k},~~~
    \bar{Z}^{I}\equiv \frac{1}{N_{s,\text{mit}}}\sum^{N_{s,\text{mit}}}_{k=1}z^{I}_{k},
\end{align}
whose averages over all possible measurement outcomes are given by \cite{PhysRevX.11.041036}
\begin{align}
    &\left<\bar{Z}^{A}\right>=\lambda^{n}\langle \psi \vert \hat{A} \vert \psi \rangle + (1-\lambda)^{n}\sum_{k=1}p_{k}^{n} \langle \psi_{k} \vert \hat{A} \vert  \psi_{k} \rangle,\\
    &\left<\bar{Z}^{I}\right>=\lambda^{n} + (1-\lambda)^{n}\sum_{k=1}p_{k}^{n}. \label{zbari}
\end{align}
One can find that the average of $\bar{Z}^{A}/\bar{Z}^{I}$ can be approximated by
\begin{align}
    \left<\frac{\bar{Z}^{A}}{\bar{Z}^{I}}\right>
    &\approx \frac{\Tr\left[\hat{A}\hat{\rho}_{\text{e}}^{n}\right]}{\Tr\left[\hat{\rho}_{\text{e}}^{n}\right]}
    \equiv\Tr\left[\hat{A}\hat{\rho}_{\text{mit}}(\phi+\phi_{0})\right]\\
    & = x_{\text{mit}} + y_{\text{mit}}\phi + O(\phi^{2}),
\end{align}
 
where $\hat{\rho}^{n}_{\text{e}}$ is the $n$th power of $\hat{\rho}_{\text{e}}$ and
\begin{align}
    \hat{\rho}_{\text{mit}}
    &\equiv \lambda'\dyad{\psi} + (1-\lambda')\sum_{k=1}p'_{k}\dyad{\psi_{k}}, \label{rhomit}
\end{align}
with
\begin{align}
    \lambda' \equiv \frac{1}{1 + \left(\frac{1-\lambda}{\lambda}\right)^{n}\sum_{k=1}p_{k}^{n} },~~~ p'_{k} \equiv \frac{p_{k}^{n}}{\sum_{l=1}p_{l}^{n}}.
\end{align}
Note that $\lambda'$ is the dominant eigenvalue of a mitigated state $\hat{\rho}_{\text{mit}}$.
We now set an estimator as
\begin{align}
    \phi^{\text{est}}_{\text{mit}}=\frac{1}{y_{\text{id}}}\left(\bar{A}_{\text{mit}}-x_{\text{id}}\right), \label{estimatormit}
\end{align}
where $\bar{A}_{\text{mit}} \equiv {\bar{Z}^{A}}/{\bar{Z}^{I}}$ and we again used the ideal coefficients $x_\text{id}$ and $y_\text{id}$ in Eq.~\eqref{xydef} because we do not know $x_\text{mit}$ and $y_\text{mit}$ for an unknown noise. 
$\phi^{\text{est}}_{\text{mit}}$ is then still a biased estimator, whose bias is given by (See Fig.~\ref{fig:schematic}.)
\begin{align}
    \left<\phi^{\text{est}}_{\text{mit}}\right>-\phi 
    &= \frac{x_{\text{mit}}-x_{\text{id}}+(y_{\text{mit}}-y_{\text{id}})\phi}{y_{\text{id}}}+O(\phi^{2}) \label{estiavgmit}\\
    &\equiv B_{\text{mit}}(\phi,\phi_{0},\Delta), \label{biasmit}
\end{align}
 and the corresponding MSE is \cite{PhysRevX.11.041036}
\begin{align}
    &\delta^{2}\phi_{\text{mit}}
    \approx \left(B_{\text{mit}}\right)^{2}\\
    &+\frac{1}{N_{s,\text{mit}}y^{2}_{\text{id}}}\bigg[\frac{1-\Tr\left[\hat{A}\hat{\rho}^{n}_{\text{e}}\right]^{2}}{\Tr\left[\hat{\rho}^{n}_{\text{e}}\right]^{2}} +\frac{\Tr\left[\hat{A}\hat{\rho}^{n}_{\text{e}}\right]^{2} (1-\Tr\left[\hat{\rho}^{n}_{\text{e}}\right]^{2})}{\Tr\left[\hat{\rho}^{n}_{\text{e}}\right]^{4}} \bigg]. \label{esterrmit}
\end{align}
In Eq.~\eqref{esterrmit}, the first term $(B_{\text{mit}})^2$ is the bias error and the second term is the statistical error. We consider $2nN_{s,\text{mit}}$ number of samples for the mitigation case. For a fair comparison between the error case and the mitigation case, we will exploit the same number of samples for an estimation, such that $N_{s}=2nN_{s,\text{mit}}$.

To summarize, the effect of VPEM is to replace the error state $\hat{\rho}_e$ with the purified state $\hat{\rho}_\text{mit}= \hat{\rho}_e^n/\Tr\left[\hat{\rho}_e^n\right]$ for the order $n$.
Since $\lambda$ is the largest eigenvalue, the $\lambda^{n}$ term becomes dominant. In contrast, the coefficient $(1-\lambda)^{n}p_{k}^{n}$, associated with the orthogonal subspace to the dominant eigenvector, is suppressed as $n$ increases. 
Therefore, the mitigated state $\hat{\rho}_{\text{mit}}$ in Eq.~\eqref{rhomit} converges to the dominant eigenvector $\dyad{\psi}$, which results in $\text{Tr}[\hat{A}\hat{\rho}_{\text{mit}}] \approx \langle \psi \vert \hat{A} \vert \psi \rangle$. 
Hence, if $y_{\text{id}}$ is not too small and the dominant eigenvector $\ket{\psi}$ and the ideal state $\ket{\psi_{\text{id}}}$ are close enough, such that $(\langle \psi \vert \hat{A} \vert \psi \rangle-\langle \psi_{\text{id}} \vert \hat{A} \vert \psi_{\text{id}} \rangle)/y_{\text{id}} \approx 0$, the bias error of the mitigation case $(B_{\text{mit}})^{2}$ would be smaller than that of the error case $(B_{\text{e}})^{2}$. 
Fig.~\ref{fig:schematic} illustrates how VPEM reduces the bias.

\section{Crucial factors for reducing bias via VPEM} \label{S3V}
In the previous section, we observed that VPEM effectively purifies an error state to its dominant eigenvector.
In this section, we further analyze the efficiency of VPEM concerning the distance between the dominant eigenvector and the ideal state with respect to an observable. Furthermore, we show that when VPEM is applied to quantum metrology, an additional factor must be necessarily considered, which is the reference point.

\subsection{Expectation value of $\hat{A}$ over the dominant eigenvector \label{S3AV}}
We analyze how the difference of the expectation values of the observable $\hat{A}$ over the dominant eigenvector of $\hat{\rho}_e$ and the ideal state affects the efficacy of VPEM.
To this end, we find an expression of a bias in the error case:
\begin{align}
    &B_{\text{e}}(\phi,\phi_{0},\Delta)=\frac{x_{\text{e}}(\phi_{0})-x_{\text{id}}(\phi_{0})+\phi\left[y_{\text{e}}(\phi_{0})-y_{\text{id}}(\phi_{0})\right]}{y_{\text{id}}(\phi_{0})}\\
    &=\frac{1}{y_{\text{id}}(\phi_{0})}\sum_{k=1}^{n-1}\left[ f_{k}(\phi_{0}) + \phi  \pdv{f_{k}(\phi)}{\phi}\bigg\vert_{\phi=\phi_{0}}\right]\Delta^{k} + O(\Delta^{n}),\label{Bedelta}
\end{align}
where we defined
\begin{align}
    &f_{1}(\phi)\equiv a_{1}(\phi)  - \lambda_{1}(\phi)b_{0}(\phi),\\
    &f_{k}(\phi)\equiv a_{k}(\phi) + \sum_{l=1}^{k-1}\lambda_{l}(\phi)a_{k-l}(\phi) - \sum_{l=1}^{k}\lambda_{l}(\phi)b_{k-l}(\phi),
\end{align}
for $k \geq 2$. Here, $\lambda_{k}$, $a_{k}$ and $b_{k}$ are coefficients from the following expansions:
\begin{align}
    &\lambda = 1 - \sum_{k=1}^{\infty}\lambda_{k}(\phi+\phi_{0})\Delta^{k}, \label{lambdaterm}\\
    &\langle \psi \vert \hat{A} \vert \psi \rangle - \langle \psi_{\text{id}} \vert \hat{A} \vert \psi_{\text{id}} \rangle = \sum_{k=1}^{\infty}a_{k}(\phi+\phi_{0})\Delta^{k}, \label{dominant}\\ 
    &\sum_{k=1}p_{k}\langle \psi_{k}  \vert \hat{A} \vert \psi_{k}\rangle - \langle \psi_{\text{id}} \vert \hat{A} \vert \psi_{\text{id}} \rangle = \sum_{k=0}^{\infty}b_{k}(\phi+\phi_{0})\Delta^{k}.\label{bkterm}
\end{align}

Eq.~\eqref{Bedelta} shows that the bias of error case consists of three distinct terms: (i) the difference between the expectation value of $\hat{A}$ over the dominant eigenvector and the ideal state $\langle \psi \vert \hat{A} \vert \psi \rangle - \langle \psi_{\text{id}} \vert \hat{A} \vert \psi_{\text{id}} \rangle$, characterized by $a_{k}$, (ii) the difference between the dominant eigenvalue $\lambda$ and $1$, identified by $\lambda_{k}$, and (iii) the difference between the expectation value of $\hat{A}$ over the tail terms $\sum_{k=1}p_{k}\langle \psi_{k}  \vert \hat{A} \vert \psi_{k}\rangle - \langle \psi_{\text{id}} \vert \hat{A} \vert \psi_{\text{id}} \rangle$, expressed by $b_{k}$. 
After applying VPEM, the bias becomes
\begin{align}
    &B_{\text{mit}}(\phi,\phi_{0},\Delta) \nonumber \\
    &=\frac{x_{\text{mit}}(\phi_{0})-x_{\text{id}}(\phi_{0})+\phi\left[y_{\text{mit}}(\phi_{0})-y_{\text{id}}(\phi_{0})\right]}{y_{\text{id}}(\phi_{0})}  \\
    &=\frac{1}{y_{\text{id}}(\phi_{0})}\sum_{k=1}^{n-1}\left[ a_{k}(\phi_{0})+\phi\pdv{a_{k}(\phi)}{\phi}\bigg\vert_{\phi=\phi_{0}}\right]\Delta^{k}+O(\Delta^{n}). \label{Bmitdelta}
\end{align}
One can easily show Eq.~\eqref{Bmitdelta} using the definition of mitigated state $\hat{\rho}_{\text{mit}}$ in Eq.~\eqref{rhomit} and the expansions in Eqs.~\eqref{dominant}-\eqref{bkterm}. 

A crucial difference from the error case is that the bias $B_{\text{mit}}$ up to the order of $\Delta^{n-1}$ only consists of $a_{k}$ which comes from the difference between the expectation value of $\hat{A}$ over the dominant eigenvector and the ideal state $\langle \psi \vert \hat{A} \vert \psi \rangle - \langle \psi_{\text{id}} \vert \hat{A} \vert \psi_{\text{id}} \rangle$. 
Therefore, $\langle \psi \vert \hat{A} \vert \psi \rangle - \langle \psi_{\text{id}} \vert \hat{A} \vert \psi_{\text{id}} \rangle$ dictates the leading order in $\Delta$ of the bias while other contributions from $b_k$ and $\lambda_k$ are suppressed by VPEM.
As a consequence, if $\langle \psi \vert \hat{A} \vert \psi \rangle - \langle \psi_{\text{id}} \vert \hat{A} \vert \psi_{\text{id}} \rangle=0$, $B_{\text{mit}}=O(\Delta^{n})$ while $B_{\text{e}}=O(\Delta)$, which clearly shows that in the small regime of $\Delta$, the bias can be reduced by applying VPEM. 
However, if the dominant eigenvector and the ideal state are not close enough, especially when $a_{1}(\phi_{0})\neq 0$, we emphasize that VPEM cannot even guarantee the constant factor reduction of the bias even for small noise. 
More specifically, for the case when $a_{1}(\phi_{0})\neq 0$, the leading orders of the biases are given by
\begin{align}
    \abs{B_{\text{e}}} &\approx \abs{\frac{ a_{1}(\phi_{0})  - \lambda_{1}(\phi)b_{0}(\phi_{0})}{y_{\text{id}}(\phi_{0})}}\Delta,\\
    \abs{B_{\text{mit}}} &\approx \abs{\frac{ a_{1}(\phi_{0})}{y_{\text{id}}(\phi_{0})}}\Delta.
\end{align}
Note that since $\lambda$ is the eigenvalue of the error state, it should be less than $1$ for any given $\Delta$, which means the $\lambda_{1}$ in Eq.~\eqref{lambdaterm} should be positive for small $\Delta$. 
In contrast, we emphasize that $a_{1}$ and $b_{0}$, in Eqs.~\eqref{dominant},~\eqref{bkterm}, can be either positive or negative; as a result, there could be a case where $\abs{B_{\text{e}}}<\abs{B_{\text{mit}}}$. 
Especially, in Sec.~\ref{prodofcssimul}, we study the case where the magnitude of a bias increases even if one applies VPEM.


We note that there have been analyses about the efficiency of VPEM~\cite{PhysRevX.11.031057,PhysRevX.11.041036,koczor2021dominant}. However, their analysis may not be directly applicable to VPEM-assisted quantum metrology. In Refs.~\cite{PhysRevX.11.031057,PhysRevX.11.041036,koczor2021dominant}, they showed that the efficacy of VPEM is often guaranteed in practice,  mainly focusing on random states, while, quantum metrology often exploits highly structured states.

\subsection{Reference point of the parameter} \label{S3BV}

\begin{figure}[t]
    \centering
    \includegraphics[width=0.95\linewidth]{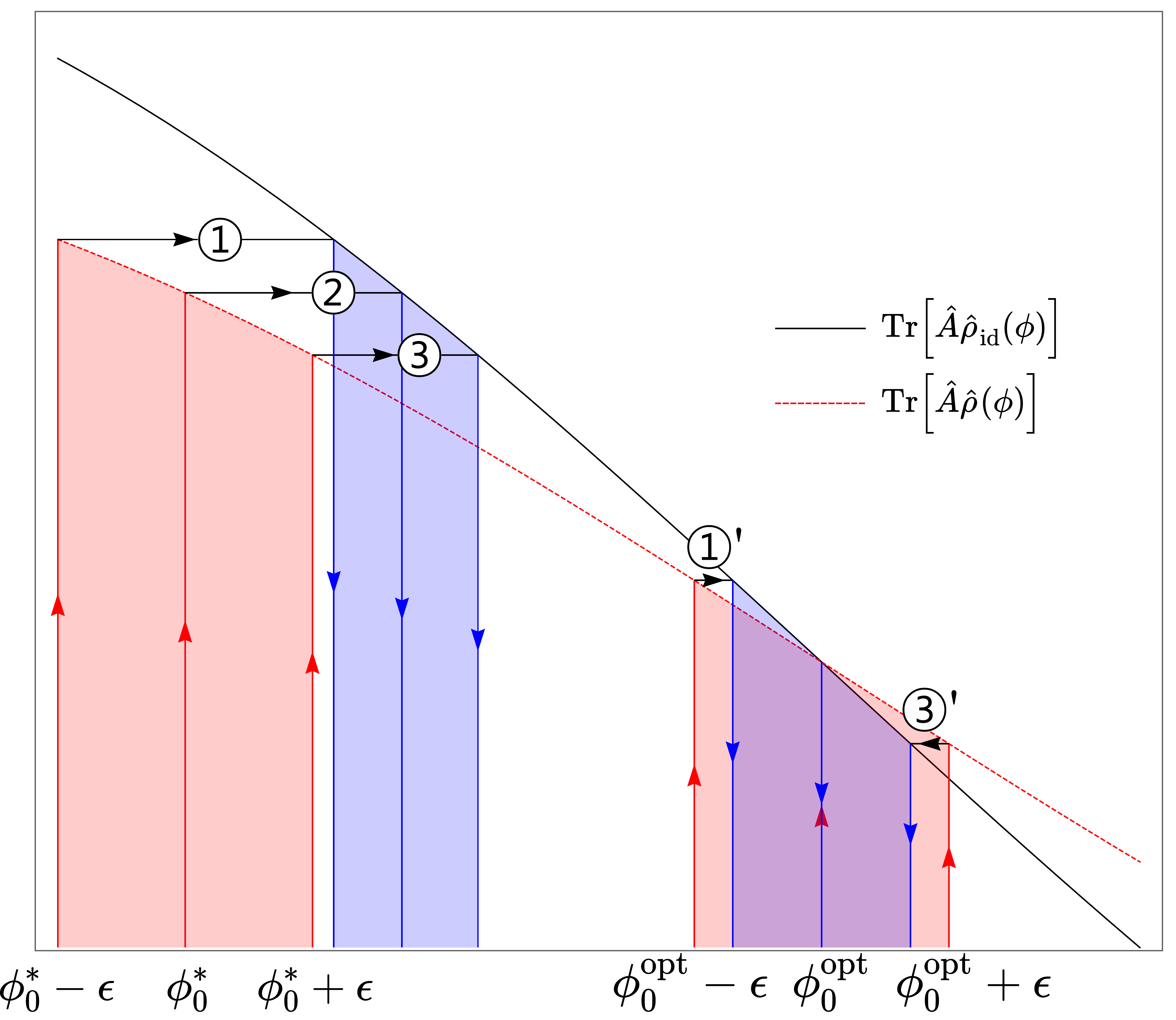}
    \caption{Schematic diagram of how the reference point affects the bias. Here $\hat{\rho}$ indicates $\hat{\rho}_{\text{e}/\text{mit}}$. We consider local parameter estimation where the parameter is in the range $\left[-\epsilon,\epsilon \right]$. The left red \& blue-colored regions correspond to the estimation when the reference point is $\phi^{*}_{0}$ which is not an optimal reference point. The three horizontal black right-headed arrows \circlenum{1}, \circlenum{2} and \circlenum{3} are the bias when the true parameter $\phi$ is $-\epsilon$, $0$ and $\epsilon$ each. In contrast, the right red and blue-colored regions correspond to the estimation when the optimal reference $\phi^{\text{opt}}_{0}$ is chosen. When the parameter is $0$, there is no bias. The two horizontal black arrows \circlenum{1}' and \circlenum{3}' are the bias when the parameter is $-\epsilon$ and $\epsilon$ each. One can find that \circlenum{1}' and \circlenum{3}' are shorter than \circlenum{1}, \circlenum{2} and \circlenum{3}, which shows reduced bias by optimal reference point.
    }
    \label{fig-referencepoint}
\end{figure}

We now introduce another important factor that plays a crucial role in VPEM for quantum metrology, which is the reference point.
Recall that, according to Eqs.~\eqref{biase} and \eqref{biasmit}, the bias of both error case and mitigation case can be expressed as
\begin{align} \label{biasterms}
    B(\phi,\phi_{0},\Delta)
    &\approx \frac{x(\phi_{0})-x_{\text{id}}(\phi_{0})+\left[y(\phi_{0})-y_{\text{id}}(\phi_{0})\right]\phi}{y_{\text{id}}(\phi_{0})}
\end{align}
Here $B(\phi,\phi_{0},\Delta)$, $x$, $y$ and $\hat{\rho}$ stand for $B_{\text{e/mit}}$, $x_{\text{e/mit}}$, $y_{\text{e/mit}}$ and $\hat{\rho}_{\text{e/mit}}$. We note that reference points for a mitigation case and an error case could be different. 
Because both biases depend on the reference point $\phi_{0}$, one might be able to reduce the bias by simply controlling $\phi_{0}$. 
In other words, there might be regions or points of $\phi_{0}$ that give a smaller bias than others. (See Fig. \ref{fig-referencepoint}.) Therefore, one should consider a reference point when adopting error mitigation to quantum metrology, unlike expectation value estimation~\cite{PhysRevX.11.031057, PhysRevX.11.041036}.
Otherwise, even if $\langle \psi \vert \hat{A} \vert \psi \rangle = \langle \psi_{\text{id}} \vert \hat{A} \vert \psi_{\text{id}} \rangle$, $(B_{\text{mit}})^{2}$ can be larger than $(B_{\text{e}})^{2}$ because of an inadequate reference point instead of the failure of error mitigation. We will show corresponding examples in Sec. \ref{S4V}.
Here, we suggest several ways to choose a \textit{good} reference point that gives a smaller bias than others to study the importance of a reference point. The reference points that we suggest below are examples that effectively show the importance of a reference point when one applies VPEM to quantum metrology.

First, we can consider a reference point that minimizes the zeroth order of $\phi$ of a bias.
Note that Eq.~\eqref{biasterms} consists of the zeroth order of $\phi$ which is $(x-x_{\text{id}})/y_{\text{id}}$ and the first order of $\phi$ which is $(y-y_{\text{id}})/y_{\text{id}}$. 
Since we are considering small $\phi$, one can alleviate the bias by choosing the reference point $\phi_{0}$ such that
\begin{align}
    \phi^{\text{opt}}_{0}(\Delta_{0}) = \arg~\min_{\phi_{0}}\left[ \frac{x(\phi_{0})-x_{\text{id}}(\phi_{0})}{y_{\text{id}}(\phi_{0})}\right]^{2}. \label{referencepoint}
\end{align}
Especially, if there exists $\phi^{\text{opt}}_{0}$ that makes $(x(\phi^{\text{opt}}_{0})-x_{\text{id}}(\phi^{\text{opt}}_{0}))/y_{\text{id}}(\phi^{\text{opt}}_{0})=0$, the zeroth order of $\phi$ in Eq.~\eqref{biasterms} vanishes and the bias only contains $\phi$ order, equivalently, the bias error only has the $\phi^{2}$ order term. When the $\phi^{\text{opt}}_{0}(\Delta_{0})$ in Eq.~\eqref{referencepoint} does not depend on $\Delta$, one can choose $\phi^{\text{opt}}_{0}(\Delta_{0})$ as a reference point regardless of noise strength. However, considering more general cases, the $\phi^{\text{opt}}_{0}(\Delta_{0})$ depends on noise strength.
In that case, a possible and practical choice of a \textit{good} reference point is to consider a reference point that minimizes the average of zeroth order of the bias over the noise strength using some prior knowledge, in such a way that
\begin{align}
    \phi^{\text{opt}}_{\text{0}}=\arg \min_{\phi_{0}} \int_{\Delta_{1}}^{\Delta_{2}} p(\Delta)\left[\frac{x(\phi_{0})-x_{\text{id}}(\phi_{0})}{y_{\text{id}}(\phi_{0})}\right]^{2}d\Delta \label{averagedreferencepoint}
\end{align}
where $p(\Delta)$ is a prior distribution of noise strength $\Delta$.
For the remainder of this paper, we call both reference points that satisfy Eq. \eqref{referencepoint} and \eqref{averagedreferencepoint} as an optimal reference point (optimal in terms that it minimizes the zeroth order of $\phi$) without confusion; all the examples we suggest, clearly show which reference point that we choose.
We denote the optimal reference point for an error case as $\phi^{\text{opt}}_{0,\text{e}}$ and a mitigation case as $\phi^{\text{opt}}_{0,\text{mit}}$.

Lastly, we present three different cases and compare $B_{\text{e}}$ and $B_{\text{mit}}$ when one chooses the optimal reference points for each case. 
For simplicity, we assume that there exists optimal reference points $\phi^{\text{opt}}_{0,\text{e}}$ and $\phi^{\text{opt}}_{0,\text{mit}}$ which make the zeroth order of bias in $\phi$ vanish.

\textit{Case 1.~--} The dominant eigenvector is the same as the ideal state, which renders ${\langle \psi \vert \hat{A} \vert \psi \rangle - \langle \psi_{\text{id}} \vert \hat{A} \vert \psi_{\text{id}} \rangle=0}$. In this case, if one applies VPEM, the leading order of a bias of mitigation case $B_{\text{mit}}$ in the noise strength is $\Delta^{n}$. In addition, by choosing $\phi^{\text{opt}}_{0,\text{mit}}$ as a reference point, $B_{\text{mit}}$ can be reduced up to $O(\phi\Delta^{n})$ which shows the efficacy of VPEM. This case corresponds to the phase estimation in the interferometer system with N00N state in the presence of phase diffusion or photon loss, studied in Sec.~\ref{phaseestN00N}, and with the product of coherent and squeezed state as a quantum probe in the presence of additive Gaussian noise which is studied in Sec.~\ref{prodofcssimul}. 

\textit{Case 2.~--} The dominant eigenvector is not close enough to the ideal state such that ${\langle \psi \vert \hat{A} \vert \psi \rangle - \langle \psi_{\text{id}} \vert \hat{A} \vert \psi_{\text{id}} \rangle=O(\phi\Delta)}$. In this case, $B_{\text{e}}$ and $B_{\text{mit}}$ have the same order of bias $O(\phi\Delta)$, i.e., VPEM cannot effectively reduce the bias. The phase estimation exploiting the product of a coherent and squeezed state which suffers from photon loss corresponds to \textit{Case 2}, which is investigated in Sec.~\ref{prodofcssimul}.

\textit{Case 3.~--} $\langle \psi \vert \hat{A} \vert \psi \rangle - \langle \psi_{\text{id}} \vert \hat{A} \vert \psi_{\text{id}} \rangle=O(\phi\Delta^{n'})$ for $n'\geq 2$. In this case, a bias cannot be suppressed beyond $O(\phi\Delta^{n'})$, no matter how large the order of VPEM $n$ is. Hence, choosing $n>n'$ does not further reduce the leading order of $\Delta$ in the bias.

A previous study of phase estimation exploiting GHZ state in the presence of amplitude damping noise \cite{yamamoto2021error} can be classified as \textit{Case 1}, where the dominant eigenvector is the same as the ideal state; therefore, the bias of mitigation case $B_{\text{mit}}$ is the order of $\Delta^{n}$, where one can benefit from error mitigation. However, we emphasize that this is a special case of our analysis, and we will investigate other cases in the following sections. In addition, we also note that there is a study that is relevant to our \textit{Case 3}. In Ref. \cite{PhysRevA.107.022608}, they study a case that shows increasing mitigation order higher than $n=2$ does not offer a further enhancement.


\section{Numerical simulation of VPEM for phase estimation scheme}\label{S4V}
In this section, we apply VPEM to phase estimation in an interferometer, which is one of the most extensively studied quantum metrological tasks. 
We exploit a N00N state or the product of a coherent and squeezed state as a quantum probe, which are well-known quantum states that enable quantum-enhanced estimation \cite{lee2002quantum,PhysRevLett.100.073601}. 
Based on our analytic study in Sec.~\ref{S3V}, we inspect whether the phase estimation scheme can benefit from VPEM in the presence of various types of noises.

\begin{figure}[t]
    \centering
    \includegraphics[width=0.95\linewidth]{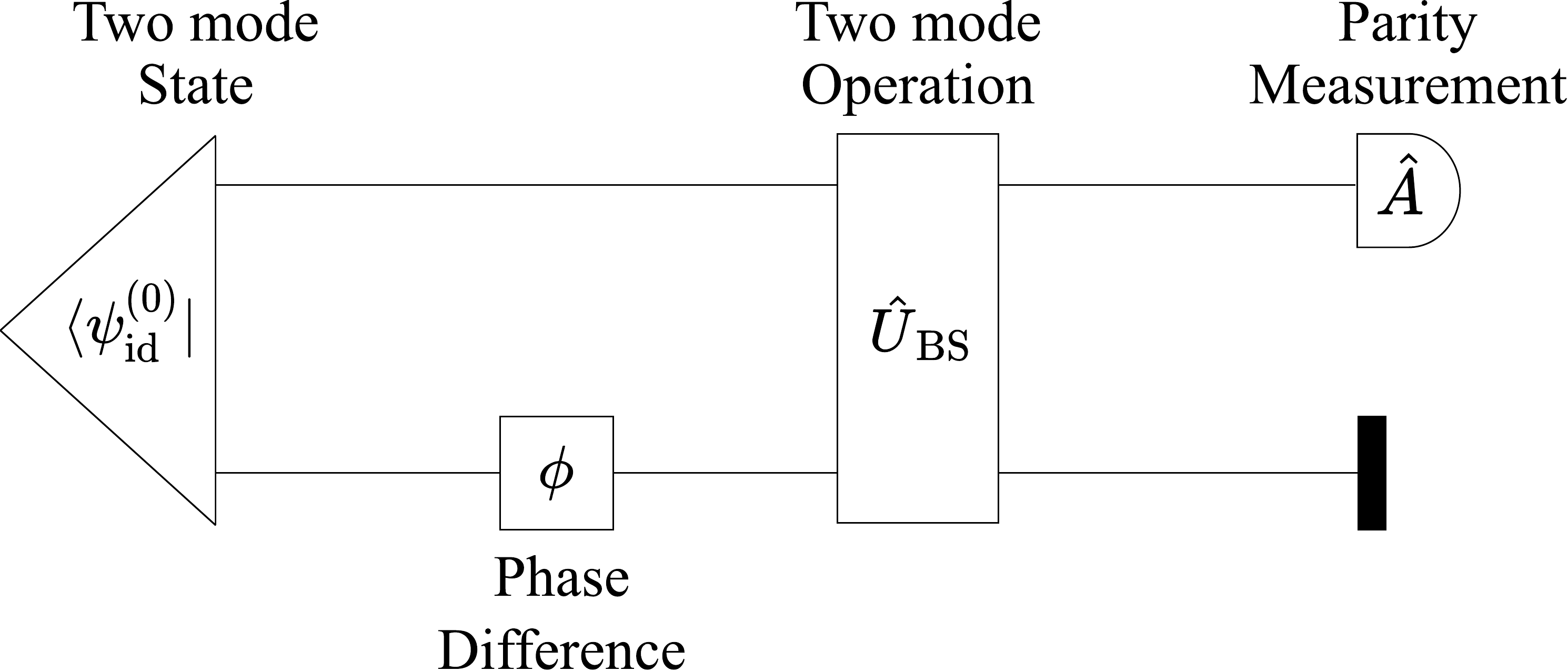}
    \caption{Interferometer system. See the main text for more details.} 
    \label{fig:Interferometer}
\end{figure}
Let us recall a phase estimation in an interferometer system, which is illustrated in Fig.~\ref{fig:Interferometer}.
First, one encodes an unknown phase $\phi$ by a phase shift operator $\hat{\Phi}(\phi)=e^{i\hat{n}_{2}\phi}$ onto a prepared two-mode quantum state $|\psi^{(0)}_{\text{id}}\rangle$, where $\hat{n}_{i}$ is the number operator in the $i$th mode. 
After a two-mode balanced beam splitter $\hat{U}_{\text{BS}}$, which renders an ideal output state $\ket{\psi_{\text{id}}(\phi+\phi_{0})} \equiv \hat{U}_{\text{BS}}\hat{\Phi}(\phi+\phi_{0})|\psi^{(0)}_{\text{id}}\rangle$, one measures the state and estimates the $\phi$ using the measurement outcomes.
Here we choose the parity operator $\hat{A}=(-1)^{\hat{n}_{1}}$ as a measurement operator.
This is a crucial property to apply the error mitigation scheme because the parity operator is both unitary and Hermitian.

In addition, the parity measurement is known to render Heisenberg scaling precision with appropriate input states \cite{hofmann2009all, anisimov2010quantum, chiruvelli2011parity,seshadreesan2011parity,PhysRevA.87.043833}.

We will consider three kinds of noise: phase diffusion, photon loss, and additive Gaussian noise which are dominant noises in phase estimation.

Phase diffusion happens when a random phase $x$ occurs during a phase $\phi$ encoding. 
Assuming that $x$ follows the Gaussian distribution, the ideal state then transforms as
\begin{align}
    \hat{\rho}_{\text{id}} \rightarrow \hat{\rho}_{\text{e}}=\frac{1}{\sqrt{2\pi\Delta} }\int_{-\infty}^{\infty}e^{-\frac{x^{2}}{2\Delta}}\hat{\rho}_{\text{id}}(\phi_{0}+\phi+x), \label{phasediffusion}
\end{align}
where $\Delta$ is a noise strength that characterizes the degree of the phase diffusion.

Photon loss transforms an annihilation operator as $\hat{a} \rightarrow \sqrt{\eta}\hat{a}+\sqrt{1-\eta}\hat{e}$, where $\hat{a}$ and $\hat{e}$ correspond to the annihilation operator for the system and the environment, respectively. 
The loss rate $0 \leq (1-\eta) \equiv \Delta \leq 1$ corresponds to noise strength. 
Since a photon-loss channel occurring on both modes with the same strength commutes with beam splitter and phase shift operation, our analysis covers a photon loss occurring from state preparation to before the measurement.

Finally, additive Gaussian noise is that a random displacement $\vec{x}=(x,p)$ occurs following Gaussian distribution, which transforms a state $\hat{\rho}$ as
\begin{align}
   \frac{1}{2\pi }\frac{1}{\sqrt{\Delta_{x}\Delta_{p}} }\int_{-\infty}^{\infty}\int_{-\infty}^{\infty}e^{-\frac{x^{2}}{2\Delta_{x}}}e^{-\frac{p^{2}}{2\Delta_{p}}}\hat{D}(\vec{x})\hat{\rho}\hat{D}^{\dagger}(\vec{x}) dx dp, \label{additivenoise}
\end{align}
where $\hat{D}(\vec{x})=e^{(x+ip)\hat{a}^{\dagger}-(x-ip)\hat{a}}$ is a displacement operator. Here $\Delta_{x}$ and $\Delta_{p}$ are noise strength. 
We consider an additive Gaussian noise occurring on the second mode of the initial quantum state $|\psi^{(0)}_{\text{id}}\rangle$.

\subsection{N00N state} \label{phaseestN00N}
\subsubsection{Ideal case}
It is known that a N00N state is one of the representative quantum probes which enables a quantum-enhanced phase estimation \cite{PhysRevA.54.R4649,lee2002quantum,Demkowicz-Dobrzanski2015}. 
In the ideal case, one exploits $|\psi^{(0)}_{\text{id}}\rangle=(\ket{N}_{1}\ket{0}_{2}+\ket{0}_{1}\ket{N}_{2})/{\sqrt{2}}$ as a quantum probe, and the expectation value of parity over $\ket{\psi_{\text{id}}(\phi+\phi_{0})}=\hat{U}_{\text{BS}}\hat{\Phi}(\phi+\phi_{0})|\psi^{(0)}_{\text{id}}\rangle$ is
\begin{align}
    \Tr\left[\hat{A}\hat{\rho}_{\text{id}}\right]=\frac{(-i)^{N}e^{{iN(\phi+\phi_{0})}}+i^{N}e^{{-iN(\phi+\phi_{0})}}}{2}.
\end{align}
Therefore, $\text{Tr}[\hat{A}\hat{\rho}_{\text{id}}]$ is either $\sin{N(\phi+\phi_{0})}$ or $\cos{N(\phi+\phi_{0})}$ with different signs. 
Since we assume that we can control the reference point, as a consequence, we can always find $\phi'_{0}$ that satisfies $\sin{N(\phi+\phi_{0})}=\cos{N(\phi+\phi'_{0})}$. Thus, without loss of generality, we consider it as $\sin{N(\phi+\phi_{0})}$. 
The MSE for the ideal case is given by
\begin{align}
    \delta^{2}\phi_{\text{id}} = \frac{1}{y_{\text{id}}^{2}}\frac{\V[\hat{A}]_{\hat{\rho}_{\text{id}}}}{N_{s}}=\frac{1}{N_{s}N^{2}},
\end{align}
which shows the Heisenberg scaling.

\begin{figure*}[t]
    \centering
    \includegraphics[width=\textwidth]{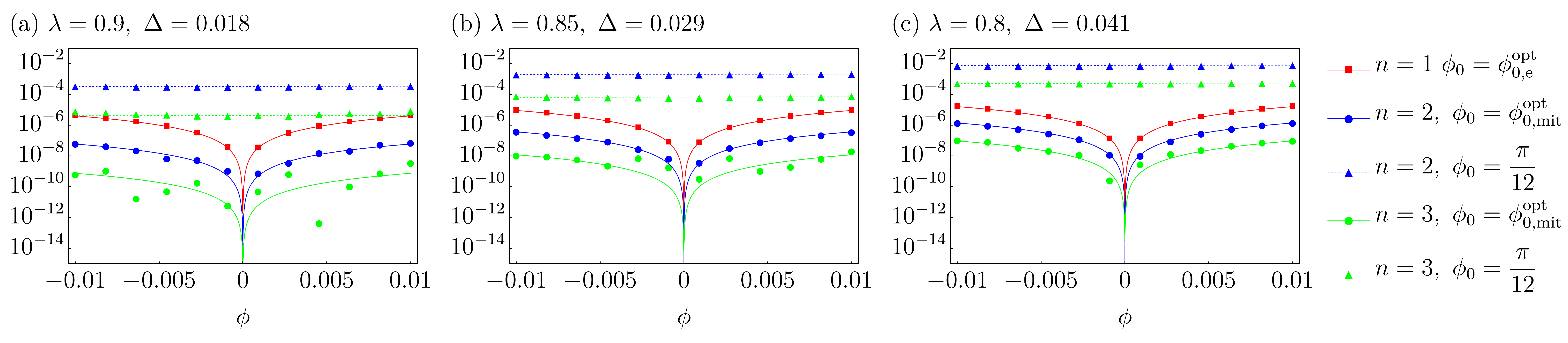}
    \caption{(a)-(c) 
    Bias errors (with log scale) of the $\phi$ estimation where $\phi \in [-0.01,0.01]$, exploiting N00N state ($N=5)$ as a quantum probe in the presence of phase diffusion with different noise strengths (namely, different dominant eigenvalues). Solid and dashed lines are theoretical values of bias errors. Circle, square, and up-triangle denote squares of the difference between estimated value and real value from simulations which are $(\phi^{\text{est}}-\phi)^{2}$ where $\phi^{\text{est}}$'s are estimators of $\phi$ (defined in Eqs. \eqref{estimatore} and \eqref{estimatormit}) with $N_{s}=10^{9}$ number of samples. We emphasize that since we exploit a large number of samples which reduces fluctuations of $\phi^{\text{est}}$'s, $(\phi^{\text{est}}-\phi)^{2}$'s can be considered as bias errors from simulation. Here we denote the error case as $n=1$ which corresponds to the red solid lines and red squares.
    } 
    \label{fig:N00N_Phase}
\end{figure*}

\subsubsection{Phase diffusion}
In the presence of phase diffusion, the ideal state transforms to an error state (which can be found using Eq. \eqref{phasediffusion}),
\begin{align}
    \hat{\rho}_{\text{e}} &=\left(\frac{1+ e^{-\frac{\Delta}{2}N^{2}}}{2}\right) \dyad{\psi_{\text{id}}} + \left(\frac{1- e^{-\frac{\Delta}{2}N^{2}}}{2}\right)\dyad{\psi_{\perp}} \\
    &\equiv \lambda \dyad{\psi_{\text{id}}} + \lambda_{\perp}\dyad{\psi_{\perp}},
\end{align}
where $\ket{\psi_{\text{id}}}=\ket{\psi_{\text{id}}(\phi+\phi_{0})}$ and $\ket{\psi_{\perp}}\equiv \hat{U}_{\text{BS}}\hat{\Phi}(\phi+\phi_{0})\left[(\ket{N}_{1}\ket{0}_{2}-\ket{0}_{1}\ket{N}_{2})/\sqrt{2}\right]$ which is orthogonal to $\ket{\psi_{\text{id}}}$. 
In this case, the dominant eigenvector is the same as the ideal state, which implies that $\langle \psi \vert \hat{A} \vert \psi \rangle = \langle \psi_{\text{id}} \vert \hat{A} \vert \psi_{\text{id}} \rangle$. 
Therefore, according to the analysis in Sec.~\ref{S3V}, one can expect that VPEM can effectively reduce the bias error. 
Let us demonstrate this by inspecting the expectation values of $\hat{A}$:
\begin{align}
    &\text{Tr}[\hat{A}\hat{\rho}_{\text{id}}]=\sin{N(\phi+\phi_{0})}\\ &=\sin{N\phi_{0}}+ (N\cos{N\phi_{0}})\phi+O(\phi^{2}),\\
    &\text{Tr}[\hat{A}\hat{\rho}_{\text{e}}]=(\lambda-\lambda_{\perp}) \sin{N(\phi+\phi_{0})}\\
    &=(\lambda-\lambda_{\perp}) \left[\sin{N\phi_{0}}+ (N\cos{N\phi_{0}})\phi\right]+O(\phi^{2}),\\
    &\text{Tr}[\hat{A}\hat{\rho}_{\text{mit}}]=\left(\frac{\lambda^{n}-\lambda^{n}_{\perp}}{\lambda^{n}+\lambda^{n}_{\perp}}\right) \sin{N(\phi+\phi_{0})}\\ &=\left(\frac{\lambda^{n}-\lambda^{n}_{\perp}}{\lambda^{n}+\lambda^{n}_{\perp}}\right) \left[\sin{N\phi_{0}}+ (N\cos{N\phi_{0}})\phi\right]+O(\phi^{2}).
\end{align}
First, we find that $\text{Tr}[\hat{A}\hat{\rho}_{\text{id}}(0)]=\text{Tr}[\hat{A}\hat{\rho}_{\text{e}}(0)]=\text{Tr}[\hat{A}\hat{\rho}_{\text{mit}}(0)]$, and $\arg \max_{\phi_{0}}y_{\text{id}}(\phi_{0})=0$, which show that the optimal reference points satisfying Eq.~\eqref{referencepoint} are $\phi^{\text{opt}}_{0,\text{e}}=\phi^{\text{opt}}_{0,\text{mit}}=0$, for both error and mitigation cases regardless of the noise strength $\Delta$ and mitigation order $n$. 
Second, once we choose the optimal reference point, the biases are given by
\begin{align}
    &B_{\text{e}} = \left(-\frac{N^{2}}{2}\Delta\right)\phi +O(\Delta^{2}), \label{biaseN00NPhase}\\
    &B_{\text{mit}} =  -2\left( \frac{ N^{2}}{4}\Delta\right)^{n}\phi+O(\Delta^{n+1}), \label{biasmitN00NPhase}
\end{align}
which shows that VPEM reduces the bias in the small range of $\Delta$ that satisfies $\frac{N^{2}}{4}\Delta \gg (\frac{N^{2}}{4}\Delta)^{n}$.

We support our results with numerical simulations. 
Fig.~\ref{fig:N00N_Phase} exhibits a simulation result of bias errors  with the N00N state in the presence of phase diffusion. We show that under the optimal reference point, VPEM can always effectively reduce the bias regardless of noise strength and the value of $\phi$. We find that $B_{\text{e}}(\phi,\phi^{\text{opt}}_{0,\text{e}},\Delta)^{2} \geq B_{\text{mit}}(\phi,\phi^{\text{opt}}_{0,\text{mit}},\Delta,n=2)^{2} \geq B_{\text{mit}}(\phi,\phi^{\text{opt}}_{0,\text{mit}},\Delta,n=3)^{2}$ as we expected through our analysis since the dominant eigenvector is equal to the ideal state. Here, we emphasize that since $\phi^{\text{opt}}_{0,\text{e}}=0$ and $\phi^{\text{opt}}_{0,\text{mit}}=0$, both biases disappear at $\phi=0$. In addition, Fig.~\ref{fig:N00N_Phase} also shows the importance of the reference point. It shows that if one does not carefully choose a reference point, an error case could have a smaller bias than a mitigation case even though $\langle \psi \vert \hat{A} \vert \psi \rangle = \langle \psi_{\text{id}} \vert \hat{A} \vert \psi_{\text{id}} \rangle$.
We find that $B_{\text{e}}(\phi,\phi^{\text{opt}}_{0,\text{e}},\Delta)^{2} \leq B_{\text{mit}}(\phi,\frac{\pi}{12},\Delta,n=3)^{2} \leq B_{\text{mit}}(\phi,\frac{\pi}{12},\Delta,n=2)^{2}$ when the dominant eigenvalue in the range $0.8 \leq \lambda \leq 0.9$, regardless of $\phi$. We emphasize that $\pi/12$ is not an optimal reference point.

\begin{figure*}[t]
    \centering
    \includegraphics[width=\textwidth]{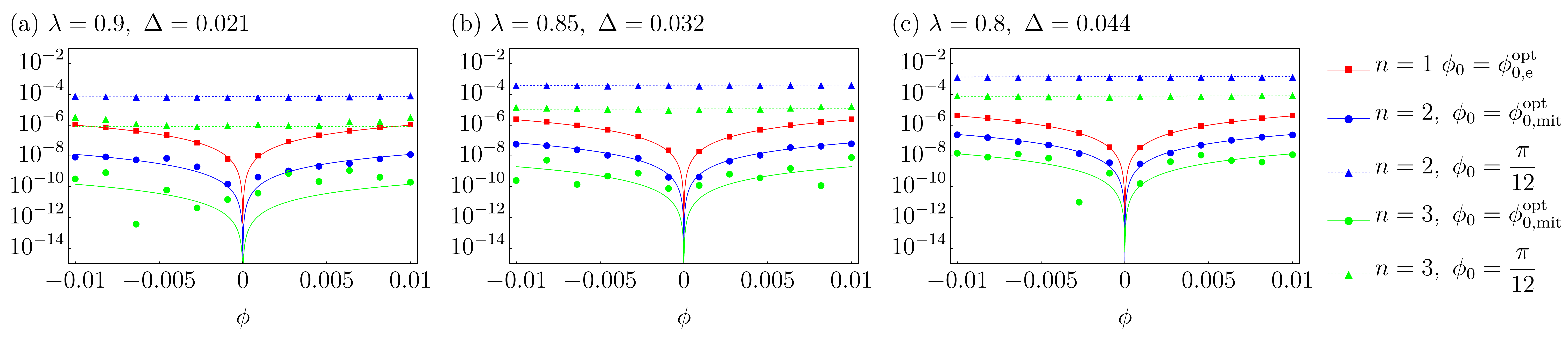}
    \caption{(a)-(c) Bias errors (with log scale) from simulation, exploiting N00N state ($N=5)$ as a quantum probe in the presence of photon loss with different noise strengths. Other features of the figure are the same as Fig. \ref{fig:N00N_Phase}.} 
    \label{fig:N00N_Loss}
\end{figure*}
\subsubsection{Photon loss}
In the presence of the photon loss, a single mode number state $\dyad{N}$ transforms to $\sum_{k=0}^{N}\binom{N}{k}\eta^{k}(1-\eta)^{N-k}\dyad{k}$, and the component $\dyad{N}{0}$ becomes $\sqrt{\eta}^{N}\dyad{N}{0}$,
where $(1-\eta)$ is the loss rate and we denote it as noise strength $\Delta$. Since we assume that photon loss occurs on each mode independently with the same noise strength and the loss channel commutes with the beam splitter and phase shift operations, one can easily find that the error state is
\begin{align}
    \hat{\rho}_{\text{e}}
    =\eta^{N} \dyad{\psi_{\text{id}}} + \left(1-\eta^{N}\right)\sum^{2N}_{k=1} p_{k} \dyad{\psi_k}.
\end{align}
Here, notice that the dominant eigenvector is equal to the ideal state. In addition,
\begin{align}
    &p_{k}=p_{N+k}=\binom{N}{k-1}\left(\frac{\eta^{k-1}(1-\eta)^{N-k+1}}{2(1-\eta^{N})}\right),\\
    &|\psi_{k}\rangle=\hat{U}_{\text{BS}}\hat{\Phi}(\phi+\phi_{0})\ket{k-1,0}_{1,2}, \\
    &|\psi_{k+N}\rangle=\hat{U}_{\text{BS}}\hat{\Phi}(\phi+\phi_{0})\ket{0,k-1}_{1,2},
\end{align}
for $k\in \{1,2,\cdots,N\}$. 
Note that $\langle \psi_{k} \vert \hat{A} \vert \psi_{k} \rangle=0$ for all $k$.
The corresponding expectation values are given by
\begin{align}
     &\text{Tr}[\hat{A}\hat{\rho}_{\text{id}}]=\sin{N(\phi+\phi_{0})}\\
     &= \sin{N\phi_{0}}+ (N\cos{N\phi_{0}})\phi+O(\phi^{2}),\\
     &\text{Tr}[\hat{A}\hat{\rho}_{\text{e}}]= (1-\Delta)^{N} \sin{N(\phi+\phi_{0})}\\
     &= (1-\Delta)^{N} \left[\sin{N\phi_{0}}+ (N\cos{N\phi_{0}})\phi\right]+O(\phi^{2}),\\
     &\text{Tr}[\hat{A}\hat{\rho}_{\text{mit}}] \nonumber \\
     &=\left(\frac{(1-\Delta)^{nN}}{\sum_{k=0}^{N}\binom{N}{k}^{n}(1-\Delta)^{nk}\Delta^{n(N-k)}}\right)\sin{N(\phi+\phi_{0})}\\
     &= \left(\frac{(1-\Delta)^{nN}}{\sum_{k=0}^{N}\binom{N}{k}^{n}(1-\Delta)^{nk}\Delta^{n(N-k)}}\right) \nonumber \\
     &~~~~\times \left[\sin{N\phi_{0}}+ (N\cos{N\phi_{0}})\phi\right]+O(\phi^{2}).
\end{align}
Again, one can find that the optimal reference points satisfying Eq.~\eqref{referencepoint} are $\phi^{\text{opt}}_{0,\text{e}}=\phi^{\text{opt}}_{0,\text{mit}}=0$. 
and the corresponding biases are
\begin{align}
    &B_{\text{e}} = \left(-N\Delta\right)\phi+O(\Delta^{2}), \\
    &B_{\text{mit}} =  -\left( N\Delta\right)^{n}\phi+O(\Delta^{3}),
\end{align}
which clearly shows that the bias is reduced by error mitigation.
We also numerically demonstrate that error mitigation can reduce bias in the presence of photon loss and the importance of a reference point, which is shown in Fig. \ref{fig:N00N_Loss}.
Similarly to the previous case, the VPEM reduces the bias error significantly because the dominant eigenvector is the same as the ideal state. However, if one does not carefully choose a reference point, the bias error of the mitigation case can be larger than the error case.

\subsection{Product of coherent and squeezed states}\label{prodofcssimul}
\subsubsection{Ideal case}
We prepare a coherent state $\ket{\alpha_{0}}_{1}$ and a squeezed vacuum state $\ket{r,0}_{2}$, and inject the states into a beam splitter which makes the entanglement between the modes. 
We exploit the state $\hat{U}_{\text{BS}}\ket{\alpha_{0}}_{1}\ket{r,0}_{2}$ as a quantum probe $|\psi^{(0)}_{\text{id}}\rangle$ which is a two-mode input state in Figure~\ref{fig:Interferometer}, the ideal state is then ${\ket{\psi_{\text{id}}}=\hat{U}_{\text{BS}}\hat{\Phi}(\phi+\phi_{0})|\psi^{(0)}_{\text{id}}\rangle}$. 
For the product of coherent state and squeezed state case, we study the effectiveness of error mitigation in the presence of photon loss and additive Gaussian noise. 
We consider $N_{c}=N_{r}=2.5$.
Unlike the previous case, the optimal reference point varies with the noise strength $\Delta$. 
To choose an appropriate reference point, we set $p(\Delta)$ in Eq.~\eqref{averagedreferencepoint} to be uniform $1/(\Delta_{2}-\Delta_{1})$, in the range $[\Delta_{1},\Delta_{2}]$
such that $\lambda(\Delta_{1})=1$ and $\lambda(\Delta_{2})=0.8$.

\subsubsection{Photon loss}
\begin{figure*}[t]
    \centering
    \includegraphics[width=\textwidth]{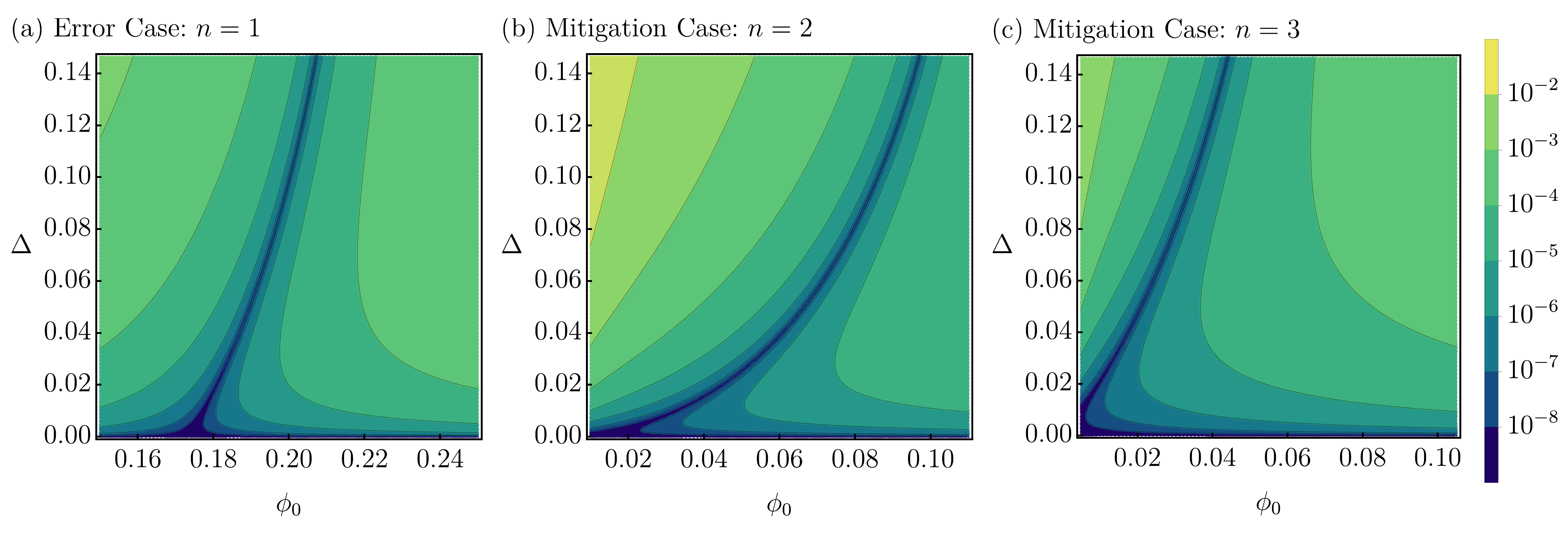}
    \caption{Contour plots of $\left[(x(\phi_{0})-x_{\text{id}}(\phi_{0}))/y_{\text{id}}(\phi_{0})\right]^{2}$ with log scale, in the presence of photon loss. Here $n=2$. (a) is for the error case and (b) is for the mitigation case. The horizontal axis is $\phi_{0}$ and the vertical axis is $\Delta$. The darkest line, where $\left[ (x(\phi_{0})-x_{\text{id}}(\phi_{0})/y_{\text{id}}(\phi_{0})\right]^{2}=0$, shows the relation between $\phi^{\text{opt}}_{0,\text{e}}$ ($\phi^{\text{opt}}_{0,\text{mit}}$) that satisfies Eq.~\eqref{referencepoint} and $\Delta$.} 
    \label{fig:contourloss}
\end{figure*}

\begin{figure*}[t]
    \centering
    \includegraphics[width=\textwidth]{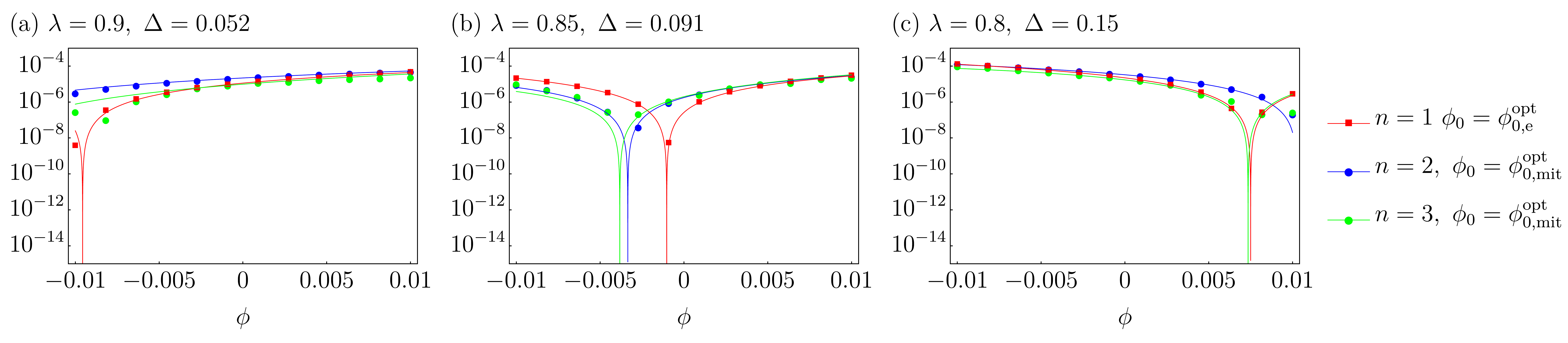}
    \caption{(a)-(c) Bias errors (with log scale) from simulation using the product of coherent state ($N_{c}=2.5$) and squeezed vacuum state ($N_{r}=2.5$) in the presence of photon loss with different noise strengths. The red-colored regions are where the bias of the error case is smaller than the mitigation one and the blue-colored regions are the opposite. Other features of the figure are the same as Fig. \ref{fig:N00N_Phase}.} 
    \label{fig:sqz_Loss}
\end{figure*}

In the presence of photon loss, the coherent state $\dyad{\alpha_{0}}$ remains as a coherent state with reduced amplitude coherent state $\dyad{\bar{\alpha_{0}}}$, where $\bar{\alpha}_{0}=\sqrt{(1-\Delta)\alpha_{0}}$. The squeezed vacuum state $\dyad{r,0}$, where the $r$ is a squeezing parameter, transforms to a squeezed thermal state \cite{serafini2017quantum} $\sum_{k=0}^{\infty} \frac{\bar{N}^{k}}{(\bar{N}+1)^{k+1}}   \dyad{\bar{r},k}_{2} $ where $\bar{N}\equiv \frac{-1+\sqrt{1-4N_{r}\{(1-\Delta)^{2}-(1-\Delta)\}}}{2}$ and $\bar{r}\equiv \frac{1}{4}\ln\left[\frac{(1-\Delta)(2N_{r}+2\sqrt{N^{2}_{r}+N_{r}})+1}{(1-\Delta)(2N_{r}-2\sqrt{N^{2}_{r}+N_{r}})+1}\right]$. Here $\ket{\bar{r},k}$ is a squeezed number state with squeezing parameter $\bar{r}$.
Again, because we assume that the photon loss occurs on both modes with the same loss rate and the loss channel commutes with linear optical operations which are phase shift and beam splitters, the error state is
\begin{align}
    \hat{\rho}_{\text{e}}=\sum_{k=0}^{\infty} \frac{\bar{N}^{k}}{(\bar{N}+1)^{k+1}} \left(\hat{\mathcal{U}} \dyad{\bar{\alpha}_{0}}_{1} \otimes  \dyad{\bar{r},k}_{2}\hat{\mathcal{U}}^{\dagger} \right),
\end{align}
where $\hat{\mathcal{U}}\equiv\hat{U}_{\text{BS}}\hat{\Phi}(\phi+\phi_{0})\hat{U}_{\text{BS}}$.

We emphasize that the dominant eigenvector
$\ket{\psi}=\hat{\mathcal{U}}\ket{\bar{\alpha}_{0}}_{1}\ket{\bar{r},k=0}_{2}$
is different from the ideal state. 
According to our analysis in Sec. \ref{S3AV}, we can expect that VPEM will not be beneficial since the dominant eigenvector and the ideal state are different. 

We numerically show the inefficacy of VPEM. For the numerical simulations, we consider the noise strength in the range $[\Delta_{1}=0,\Delta_{2}=0.146]$ and assume that $\Delta$ follows uniform distribution in the range. Here $\Delta_{2}$ satisfies $\lambda(\Delta_{2})=0.8$ where $\lambda$ is the dominant eigenvalue of the error state. The optimal reference point $\phi^{\text{opt}}_{0}$ satisfying Eq.~\eqref{referencepoint} depends on a given $\Delta$ for both error and mitigation cases. The relation between $\Delta$ and $\phi^{\text{opt}}_{0}$'s for error case, mitigation case with $n=2$ and $n=3$ are provided in Fig.~\ref{fig:contourloss}. Therefore, we consider an averaged optimal reference point that satisfies Eq. \eqref{averagedreferencepoint} which are $\phi^{\text{opt}}_{0,\text{e}}\approx0.2$, $\phi^{\text{opt}}_{0,\text{mit}}\approx0.087$ for $n=2$, and $\phi^{\text{opt}}_{0,\text{mit}}\approx0.037$ for $n=3$. Under the chosen optimal reference points, we numerically simulate the bias errors of the $\phi$ estimation exploiting the product of the coherent and the squeezed state in the presence of photon loss. Fig.~\ref{fig:sqz_Loss} shows the non-effectiveness of VPEM-based quantum metrology as implied by our analysis because of the difference between the dominant eigenvector and the ideal state.
For the most of the regions of $\phi$, $B_{\text{e}}(\phi,\phi^{\text{opt}}_{0,\text{e}},\Delta)^{2}$, $B_{\text{mit}}(\phi,\phi^{\text{opt}}_{0,\text{mit}},\Delta,n=2)^{2}$ and $B_{\text{mit}}(\phi,\phi^{\text{opt}}_{0,\text{mit}},\Delta,n=3)^{2}$ are similar in magnitudes. Moreover, there are combinations of $\Delta$ and $\phi$ where $B_{\text{e}}(\phi,\phi^{\text{opt}}_{0,\text{e}},\Delta)^{2}$ is smaller than the others.

\begin{figure}[t]
    \centering
    \includegraphics[width=\linewidth]{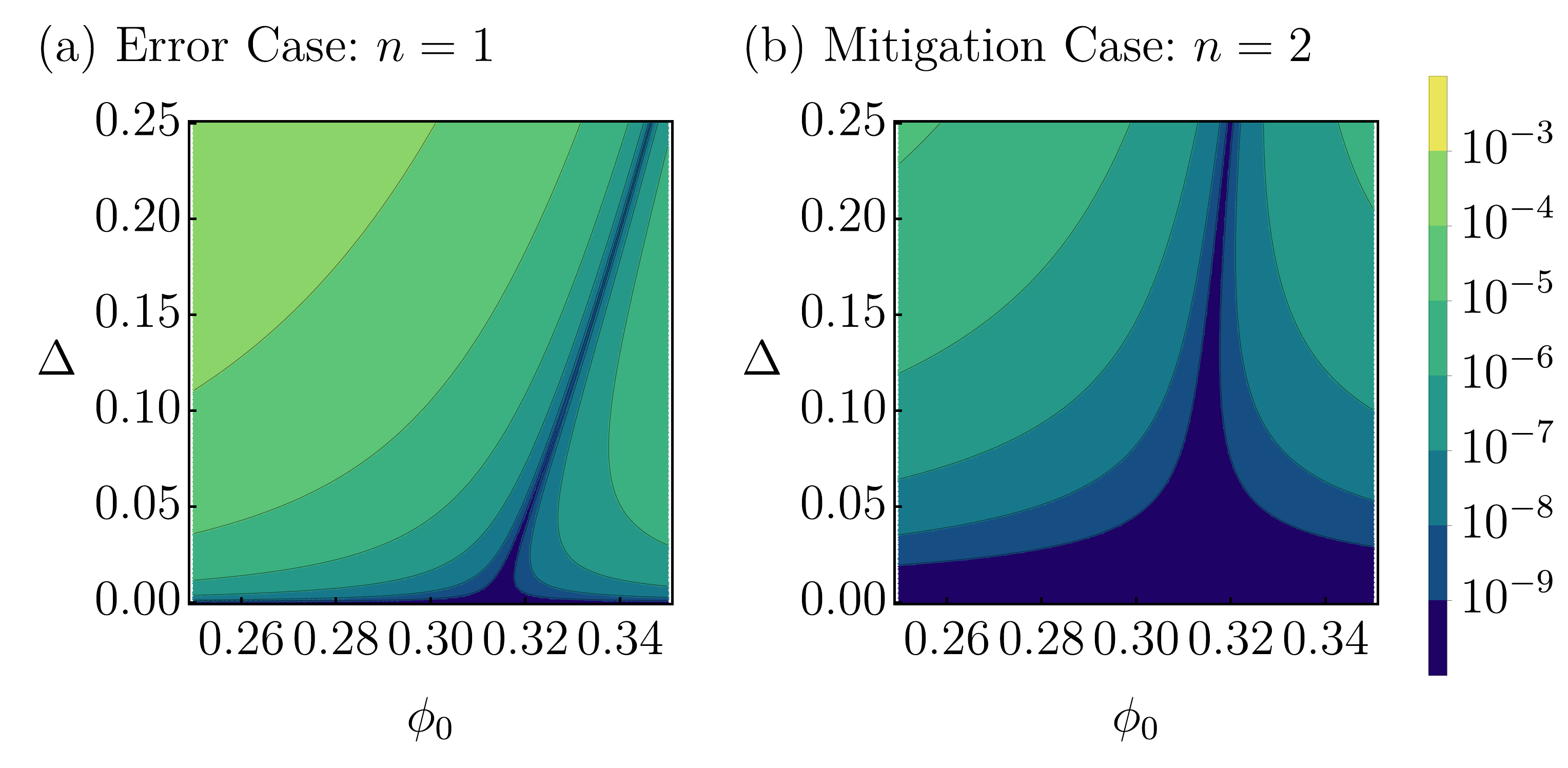}
    \caption{Contour plots of $\left[\big(x(\phi_{0})-x_{\text{id}}(\phi_{0})\big)/y_{\text{id}}(\phi_{0})\right]^{2}$ with log scale, in the presence of additive Gaussian noise. Other features of the Figure are similar to Figure \ref{fig:contourloss}.} 
    \label{fig:contourdis}
\end{figure}

\begin{figure*}[t]
    \centering
    \includegraphics[width=\textwidth]{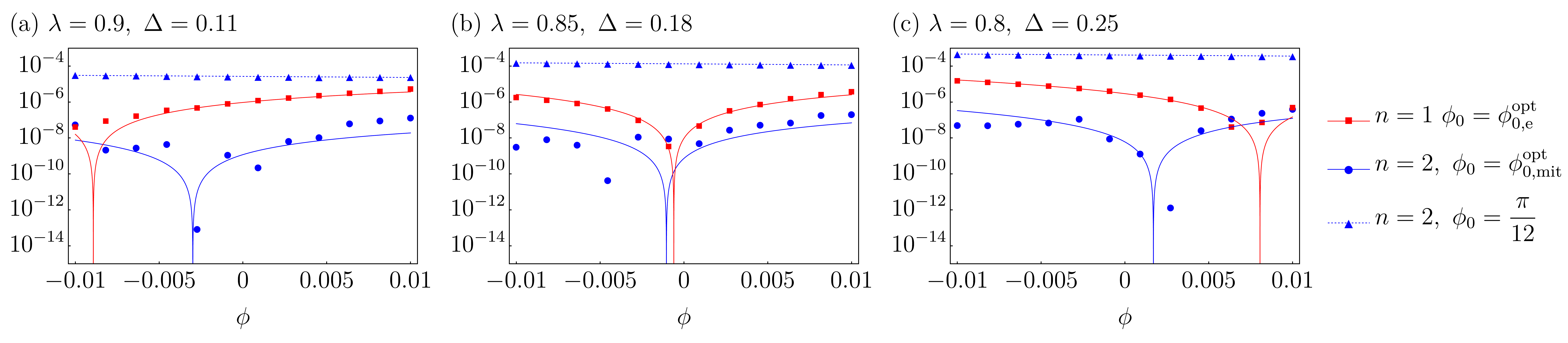}
    \caption{(a)-(c) The simulations of the bias errors (with log scale) correspond to the product of coherent state ($N_{c}=2.5$) and squeezed vacuum state ($N_{r}=2.5$) in the presence of additive Gaussian noise with different noise strengths. Other features of the figure are the same as Fig. \ref{fig:N00N_Phase}.} 
    \label{fig:sqz_Gaussian}
\end{figure*}

\subsubsection{Additive Gaussian noise}
For pedagogical purposes, we consider an additive Gaussian noise, which might not be directly relevant to experiments. 
Let us assume that the additive Gaussian noise occurs only on the second mode of the initial quantum probe $\ket{\alpha_{0}}_{1}\ket{r,0}_{2}$. 
Furthermore, we assume that $\Delta_{x}$ and $\Delta_{p}$ in Eq.~\eqref{additivenoise} are
\begin{align}
    &\Delta_{x}=\sqrt{\frac{\Delta}{2}}e^{-r},~~\Delta_{p}=\sqrt{\frac{\Delta}{2}}e^{r} \label{standarddisplace}
\end{align}
where $r$ is a squeezing parameter in $\ket{r,0}_{2}$ and $\Delta$ is a noise strength. We note that the additive Gaussian noise is a Gaussian error such that a Gaussian state remains as a Gaussian state after suffering from the noise. Similar to the photon loss case, the additive Gaussian noise makes a single-mode squeezed state to squeezed thermal state \cite{serafini2017quantum} $\sum_{k=0}^{\infty} \frac{\Delta^{k}}{(\Delta+1)^{k+1}}   \dyad{r,k}$. Since we consider the additive Gaussian noise only on the second mode of the quantum probe at the state preparation stage (before passing the first beam splitter), as a consequence, the error state is
\begin{align}
    \hat{\rho}_{\text{e}}=\sum_{k=0}^{\infty} \frac{\Delta^{k}}{(\Delta+1)^{k+1}} \bigg[\hat{\mathcal{U}} \dyad{\alpha_{0}}_{1} \otimes  \dyad{r,k}_{2}\hat{\mathcal{U}}^{\dagger} \bigg], 
\end{align}
whose dominant eigenvector is the same as the ideal state, which means $\langle \psi \vert \hat{A} \vert \psi \rangle = \langle \psi_{\text{id}} \vert \hat{A} \vert \psi_{\text{id}} \rangle$. 
In this case, $a_{1}(\phi)$ and $\pdv{a_{1}(\phi)}{\phi}$ are always $0$ regardless of $\phi$, therefore, one can expect that error mitigation can efficiently reduce bias error.

We numerically show the efficiency of VPEM. For the numerical simulations, we consider the noise strength in the range $[\Delta_{1}=0,\Delta_{2}=0.25]$ and again, assume that $p(\Delta)=1/(\Delta_{2}-\Delta_{1})$. Here $\Delta_{2}$ satisfies $\lambda(\Delta_{2})=0.8$. In an additive Gaussian noise case again, the optimal reference point that satisfies Eq. \eqref{referencepoint} depends on noise strength (See Fig. \ref{fig:contourdis}), we consider an averaged optimal reference point that satisfies Eq. \eqref{averagedreferencepoint} which are $\phi^{\text{opt}}_{0,\text{e}}\approx 0.338$ and $\phi^{\text{opt}}_{0,\text{mit}}\approx 0.318$. Figure \ref{fig:sqz_Gaussian} is a simulation result with the product of the coherent and squeezed state in the presence of additive Gaussian noise. In Fig. \ref{fig:sqz_Gaussian}, we show that even though one adopts the reference points that satisfy Eq. \eqref{averagedreferencepoint}, VPEM can effectively reduce the bias because the additive Gaussian noise let the dominant eigenvector being the ideal state. For most of $\phi$'s, one can clearly find that $B_{\text{e}}(\phi,\phi^{\text{opt}}_{0,\text{e}},\Delta)^{2} \geq B_{\text{mit}}(\phi,\phi^{\text{opt}}_{0,\text{mit}},\Delta)^{2}$. In addition, we can also find the importance of the reference point in Fig.~\ref{fig:sqz_Gaussian}. We find that $B_{\text{e}}(\phi,\phi^{\text{opt}}_{0,\text{e}},\Delta)^{2} \leq B_{\text{mit}}(\phi,\frac{\pi}{30},\Delta)^{2}$ even though $\langle \psi \vert \hat{A} \vert \psi \rangle = \langle \psi_{\text{id}} \vert \hat{A} \vert \psi_{\text{id}} \rangle$.

\section{Discussion}
We have investigated crucial factors that dictate the efficiency of VPEM when one applies the method to quantum metrology. 
We find that the dominant eigenvector has to be close to the ideal state, with respect to an observable, to reduce the bias by VPEM successfully. More specifically,  $\langle \psi \vert \hat{A} \vert \psi \rangle-\langle \psi_{\text{id}} \vert \hat{A} \vert \psi_{\text{id}} \rangle$ determines the reducible amount of bias. 
In addition, we argue that one should carefully choose an optimal reference point that minimizes bias error. Otherwise, the bias error of the mitigation case could be larger than the error case even if the dominant eigenvector is equal to the ideal state, because of the inadequate reference point. 
We emphasize that many of the practical estimation scenarios giving an additional parameter for the reference point have already been implemented experimentally, such as phase estimation schemes \cite{berni2015ab}. Based on the analysis, we analytically and numerically inspect the phase estimation scheme in the interferometer system. We consider the N00N state and the product of the coherent and the squeezed state as quantum probes in the presence of phase diffusion, photon loss, and additive Gaussian noise. We study that all the phase estimation schemes that we mentioned above can be explained by our analysis.

Finally, we point out the differences from the recent study~\cite{yamamoto2021error}.
First, while Ref.~\cite{yamamoto2021error} analyzes bias up to the zeroth order of $\phi$, we consider up to the first order of $\phi$, which is essential because  
 according to our optimal reference point strategy, the zeroth order of $\phi$ can vanish both for the error case and mitigation case. 
Second, we study the cases where VPEM cannot effectively reduce the bias, which has not been studied before, to the best of our knowledge.
Lastly, we finally present the importance of the reference point.
In the previous study, because the optimal reference point is fixed for the error case and mitigated case, the role of the reference point was not considered. 
It would be an interesting future work to find a strategy for choosing an optimal reference point other than Eq. \eqref{averagedreferencepoint}. For example, one can develop an adaptive strategy that exploits measurement outcomes to estimate a noise strength and updates a reference point using the estimated noise strength. This method may give a tailored reference point for a given noise which results in a more reduced bias than the reference point in Eq. \eqref{averagedreferencepoint}.

\section*{Acknowledgments}
H.K. and H.J. were supported by the National Research Foundation of Korea (NRF) grants funded by the Korean government (Grant Nos. NRF-2023R1A2C1006115 and NRF-2022M3K4A1097117) and by the Institute of Information \& Communications Technology Planning \& Evaluation (IITP) grant funded by the Korea government (MSIT) (IITP-2021-0-01059 and IITP-2023-2020-0-01606). 
H.K. was supported by the education and training program of the Quantum Information Research Support Center, funded through the National research foundation of Korea (NRF) by the Ministry of science and ICT (MSIT) of the Korean government (No.2021M3H3A103657313).
L.J. and C.O. acknowledge support from the ARO (W911NF-23-1-0077), ARO MURI (W911NF-21-1-0325), AFOSR MURI (FA9550-19-1-0399, FA9550-21-1-0209), AFRL (FA8649-21-P-0781), NSF (OMA-1936118, ERC-1941583, OMA-2137642), NTT Research, and the Packard Foundation (2020-71479). This work is partially supported by the U.S. Department of Energy Office of Science National Quantum Information Science Research Centers.
Y. L. acknowledges National Research Foundation of Korea a grant funded by the Ministry of Science and ICT (NRF-2022M3H3A1098237) and KIAS Individual Grant
(CG073301) at Korea Institute for Advanced Study.

\bibliography{reference.bib}
\end{document}